\documentclass{emulateapj}

\shorttitle{Magnetic Fields in M Dwarfs}
\shortauthors{Phan-Bao et al.}

\begin{document}

\title{MAGNETIC FIELD TOPOLOGY IN LOW-MASS STARS: SPECTROPOLARIMETRIC OBSERVATIONS OF M DWARFS$^{1}$}

\author{Ngoc Phan-Bao,\altaffilmark{2, 3}
Jeremy Lim,\altaffilmark{2, 4}
Jean-Fran\c{c}ois Donati,\altaffilmark{5}
Christopher M. Johns-Krull,\altaffilmark{6}
and Eduardo L. Mart\'{\i}n,\altaffilmark{7, 8}}

\altaffiltext{1}{Based on observations made at the
 Canada-France-Hawaii Telescope, operated by the National Research Council of Canada, the Centre National de la Recherche Scientifique de France and the University of Hawaii.}
\altaffiltext{2}{Institute of Astronomy and Astrophysics, Academia Sinica, P.O. Box 23-141, Taipei 106, Taiwan, ROC; pbngoc@asiaa.sinica.edu.tw; jlim@asiaa.sinica.edu.tw.}
\altaffiltext{3}{Department of Physics, HCMIU, Vietnam National University Administrative Building,
                 Block 6, Linh Trung Ward, Thu Duc District, HCM, Vietnam.}
\altaffiltext{4}{Department of Physics, Room 518, Chong Yuet Ming Physics Building,
The University of Hong Kong, Pokfulam Road, Hong Kong, China.}
\altaffiltext{5}{Laboratoire d'Astrophysique, Observatoire Midi-Pyr\'en\'ees, F-31400 Toulouse, France; donati@ast.obs-mip.fr.}
\altaffiltext{6}{Department of Physics and Astronomy, Rice University, 6100 Main Street, MS-61 Houston, TX 77005; cmj@rice.edu.} 
\altaffiltext{7}{CSIC-INTA Centro de Astrobiologia, Torrejon de Ardoz, M a d r i d, Spain; ege@iac.es.}
\altaffiltext{8}{University of Central Florida, Dept. of Physics, PO Box 162385, Orlando, FL 32816-2385.}

\begin{abstract}
The magnetic field topology plays an important role in the understanding of 
stellar magnetic activity.  While it is widely accepted that the dynamo
action present in low-mass partially convective stars (e.g., the Sun)
results in predominantly toroidal magnetic flux, the field topology in 
fully convective stars (masses below $\sim0.35$ M$_\odot$) is still under 
debate.  We report here our mapping of the magnetic field topology of the M4 dwarf G~164-31 (or Gl~490B), 
which is expected to be fully convective, based on time series data 
collected from 20 hours of observations 
spread over 3 successive nights with the ESPaDOnS spectropolarimeter.
Our tomographic imaging technique applied to time series of rotationally 
modulated circularly polarized profiles reveals an axisymmetric large-scale poloidal magnetic field on the M4 dwarf. 
We then apply a synthetic spectrum fitting technique
for measuring the average magnetic flux on the star. 
The flux measured in G~164-31 is $|Bf| = 3.2\pm0.4$~kG,
which is significantly greater than the average value of 0.68~kG determined from the imaging technique.
The difference indicates that a significant fraction of the stellar
magnetic energy is stored in small-scale structures at the surface of G~164-31.
Our H$\alpha$ emission light curve shows evidence for rotational modulation
suggesting the presence of localized structure in the chromosphere
of this M dwarf. The radius
of the M4 dwarf derived from the rotational period and the projected 
equatorial velocity is at least $30\%$ larger than that predicted from
theoretical models. We argue that this discrepancy is likely primarily
due to the young nature of G~164-31
rather than primarily due to 
magnetic field effects, indicating that age is an important
factor which should be considered in the interpretation of this observational result.
We also report here our polarimetric observations of five 
other M dwarfs with spectral types from M0 to M4.5, 
three of them showing strong Zeeman signatures.
\end{abstract}

\keywords{stars: low mass --- stars: magnetic fields --- 
techniques: spectroscopic --- techniques: polarimetric --- 
stars: individual (G 164-31, Gl 890, LHS 473, KP Tau, Gl 896B, 2E 4498)}

\section{INTRODUCTION}

Magnetic fields in the Sun and partially convective 
stars (e.g., G, K stars and possibly early M dwarfs) are produced by the so-called $\alpha\Omega$
dynamo mechanism operating at the interface between the convective envelope and the
radiative core, where differential rotation is strongest, known as the tachocline. 
The action of differential rotation on a weak poloidal field
at the base of the convective zone results in a large-scale and predominantly 
toroidal sub-surface field produced by this dynamo action \citep{parker75}.
However, when this toroidal field emerges through the surface of the Sun,
the magnetic flux tubes are oriented in a primarially radial direction 
in the photosphere (e.g. see the review by Solanki 2009).

Stars with masses below $\sim0.35$~M$_{\odot}$ ($\sim$M3)
are expected to be fully convective as suggested by standard
models (e.g., \citealt{chabrier97}). This limit probably
shifts toward lower masses due to the influence
of the magnetic field \citep{mullan,chabrier07}.
In fully convective stars, 
the lack of a radiative core is expected to preclude
the $\alpha\Omega$ dynamo, maintained
by the combined action of differential rotation ($\Omega$ effect)
and cyclonic convection ($\alpha$ effect). This raises the question
of what type of dynamo produces strong magnetic activity as observed
in mid to late M and brown dwarfs from 
photospheric Zeeman splitting \citep{jk96,rn07} to
chromospheric H$\alpha$ and Ca~{\small II}~IRT \citep{liebert,pb06,schmidt07},
coronal X-ray \citep{audard,stelzer06,robrade}, and radio \citep{berger02,pb07,berger08,hallinan08,osten09}
emissions.

An alternative dynamo known as the
so-called $\alpha^{2}$ dynamo maintained by convection alone
was proposed \citep{roberts72}. It has been developed 
for fully convective pre-main sequence stars 
\citep{schu,rudiger,kuker97,kuker99} and late M dwarfs and brown dwarfs 
\citep{dobler,chabrier06,browning}.
In general, these models predict large-scale fields, and magnetic field
detections are now known for a number of active objects in these groups 
\citep{saar94,jk96,donati06a,morin08,berger09}.
However, the models disagree with each other
on the magnetic field morphology and they disagree with the results from
some phase-resolved spectropolarimetric observations \citep{donati06a,morin08}.
For example, the latest dynamo simulations \citep{browning} 
sucessfully predict that the magnetic fields possess 
large-scale fields with a lack of differential rotation
in rapidly rotating M dwarfs. 
However, the simulations predict the field morphology
in fully convective stars should be mostly toroidal, whereas results of 
\citet{donati06a} and \citet{morin08} show dominantly
poloidal fields in 5 M3-M4.5 dwarfs.

In this paper, we present our mapping of
the M4 dwarf G~164-31 based on 20 hours of spectropolarimetric observations 
spread over 3 successive nights.
To date, there are only two M4 dwarfs (V374 Peg, \citealt{donati06a};
G~164-31, this paper) that have been densely monitored 
over only a few rotation cycles so as to preclude intrinsically 
long-term magnetic variability, which can affect the mapping result. 
In addition, for the first time
a synthetic spectrum fitting technique \citep{marcy82,saar88,jk96} 
is applied for measuring the mean field strength and the filling factor in G~164-31.
Together, the polarimetric field mapping and the mean field measurements
provide us a better accounting of both large- and small-scale fields in 
one of these two rapid rotators.
We also present our Zeeman signature search in Stokes V for 
5 other M-dwarfs with spectral types
ranging from M0 to M4.5: 
Gl 890 (M0V), LHS 473 (M2.5V), KP Tau (M3V), Gl 896B (M4.5V) and 2E~4498 (M4.5V).
Strong circular polarization is detected in KP Tau, Gl 896B and 2E~4498.

We present our targets in \S~2 and the observation and the data reduction in \S~3.
We apply the tomographic imaging technique for constructing 
the field topology of Gl~490B in \S~4. 
We discuss the field topology of the 6 M dwarfs in \S~5 and summarize the results 
in \S~6.

\section{TARGETS}
\subsection{Targets}
All the 6 targets have spectral types from M0V to M4.5V and their physical parameters are listed
in Table~\ref{tab_Bz}. From our mass estimates (see Table~\ref{tab_Bz} and references therein), 
three M dwarfs (KP~Tau, G~164-31 and 2E~4498) have masses estimated
below 0.35~M$_{\odot}$
(or absolute magnitudes fainter than $M_{\rm K}=6.62$) and
are expected to be fully convective according to
standard models. Two early-M dwarfs (Gl 890 and LHS 473) are partially convective.
The absolute magnitude of the unresolved binary system Gl 896Bab is $M_{\rm K}=7.28$, which gives
$M_{\rm K}>7.28$ for each companion.  Both components are therefore expected to be
fully convective. 
\subsubsection{Gl 890 (HK Aqr), M0V}
This flaring M0 dwarf is a rapid rotator with $v$~sin$i=70$~km~s$^{-1}$
\citep{young84}. A photometric period of 0.431 days was determined
by \citet{young90}. The star has been well studied for magnetic activity, but no circular
polarization observations have been reported so far.
\citet{rao} detected a steady soft X-ray emission
from the source, indicating coronal activity. 
Its X-ray luminosity (0.2-4.5 keV) was measured by XMM-Newton at
log$L_{\rm X}$(ergs~s$^{-1}$) $\approx$ 29.0 \citep{xmm}.
A large flarelike event was also detected by \citet{singh}
from the X-ray light curve. Gl 890 was observed with the 
Australia Telescope Compact Array (ATCA) at 6~cm over the
full rotation phase in 1990, but the source was not detected 
at an upper flux limit of $\approx$0.3~mJy \citep{lim}.
It was subsequently detected as a weak (below 0.3~mJy) 
slowly varying source with the Very Large Array (VLA)
(\citealt{lim}, quoting a private communication by M. G\"udel \& A. D. Benz 1992).

\subsubsection{LHS 473 (GJ 752A), M2.5V}
LHS 473 is an M2.5 dwarf, a companion of VB 10 (M8), and has
a low rotational velocity $v$~sin$i < 2.6$~km~s$^{-1}$ \citep{del98}.
\citet{vaiana} reported a soft X-ray (0.2-4.0 keV) detection with 
log$L_{\rm X}=27.11$.
\citet{leto} carried out VLA radio observations 
but the star was not detected in all 4 different observed radio bands. These observations 
indicate a low coronal activity level in LHS 473.
\citet{stauffer86} also reported weak chromospheric activity
with an H$\alpha$ absorption equivalent width of 0.35~\AA \citep{cram87}.
\citet{jk96} classified LHS 473 an inactive M dwarf.

\subsubsection{KP Tau (Wolf 1246, RX J0339.4$+$2457), M3V}
This M3 dwarf, in contrast with LHS 473, is a rapid rotator with
$v$~sin$i=32$~km~s$^{-1}$ \citep{moch}.
The dwarf is a ROSAT source (0.1-2.4 keV) with
log$L_{\rm X}=28.13$ \citep{fleming98,huensch}, implying that
the coronal activity level in KP~Tau is over an order of magnitude
higher than that in LHS 473 (which has a similar spectral type).
To date, no radio observations of the star have been reported.

\subsubsection{G 164-31 (Gl 490B), M4V} 
G 164-31 is a star with a spectral type of M4V \citep{reid95} at a
distance of 18.1~pc \citep{esa}, which is
the companion of G 164-32 (Gl 490A, M0.5V) at a large angular separation of 15$\arcsec$.
At the given distance and its apparent 2MASS $K$-band magnitude of
$K = 8.02$, we derive its absolute
magnitude $M_{\rm K}=6.73$. 
This fast rotator with $v$sin$i=34$~km~s$^{-1}$
exhibits high coronal activity in X-ray emission (0.3-3.5 keV) with
log$L_{\rm X}=29.05$ \citep{fleming89}.
No radio observations of the star have been reported to date.
Using the \citet{segransan} empirical relationship of $M_{\rm K}$ versus radius, we derive
$R_{\star}=0.34$~R$_{\odot}$, which is in good agreement with 
theoretical models (e.g., \citealt{b98}) for inactive dwarfs (see \citealt{demory}).
This estimated radius yields a maximum rotational period of $\sim$12.1~hours. 

\subsubsection{Gl 896B (EQ Peg B), M4.5V}
The Gl 896 system is a quadruple Gl 896AabBab \citep{del99,oppen}. 
Gl 896 Bab is a single-lined spectroscopic binary with a joint spectral type
of M4.5 and a joint $v$~sin$i$ of 24.2~km~s$^{-1}$ \citep{del98}.
Gl 896Bab was detected at radio frequencies \citep{gudel93} of 4.9~GHz (6 cm)
and 8.5~GHz (3.6 cm). 
The joint X-ray luminosity (0.1-2.4 keV) of the Gl 896 AabBab system is 
log$L_{\rm X}=28.84$ \citep{gudel93}.
\citet{robrade} reported Gl~896Bab was on average a factor of 3.5
weaker than Gl~896Aab, we therefore derive log$L_{\rm X}=28.19$
for Gl~896Bab.

\subsubsection{2E 4498 (RX J2137.6$+$0137), M4.5V} 
\begin{figure}
\vskip 1in
\hskip -0.25in
\centerline{\includegraphics[width=7.0in,angle=-90]{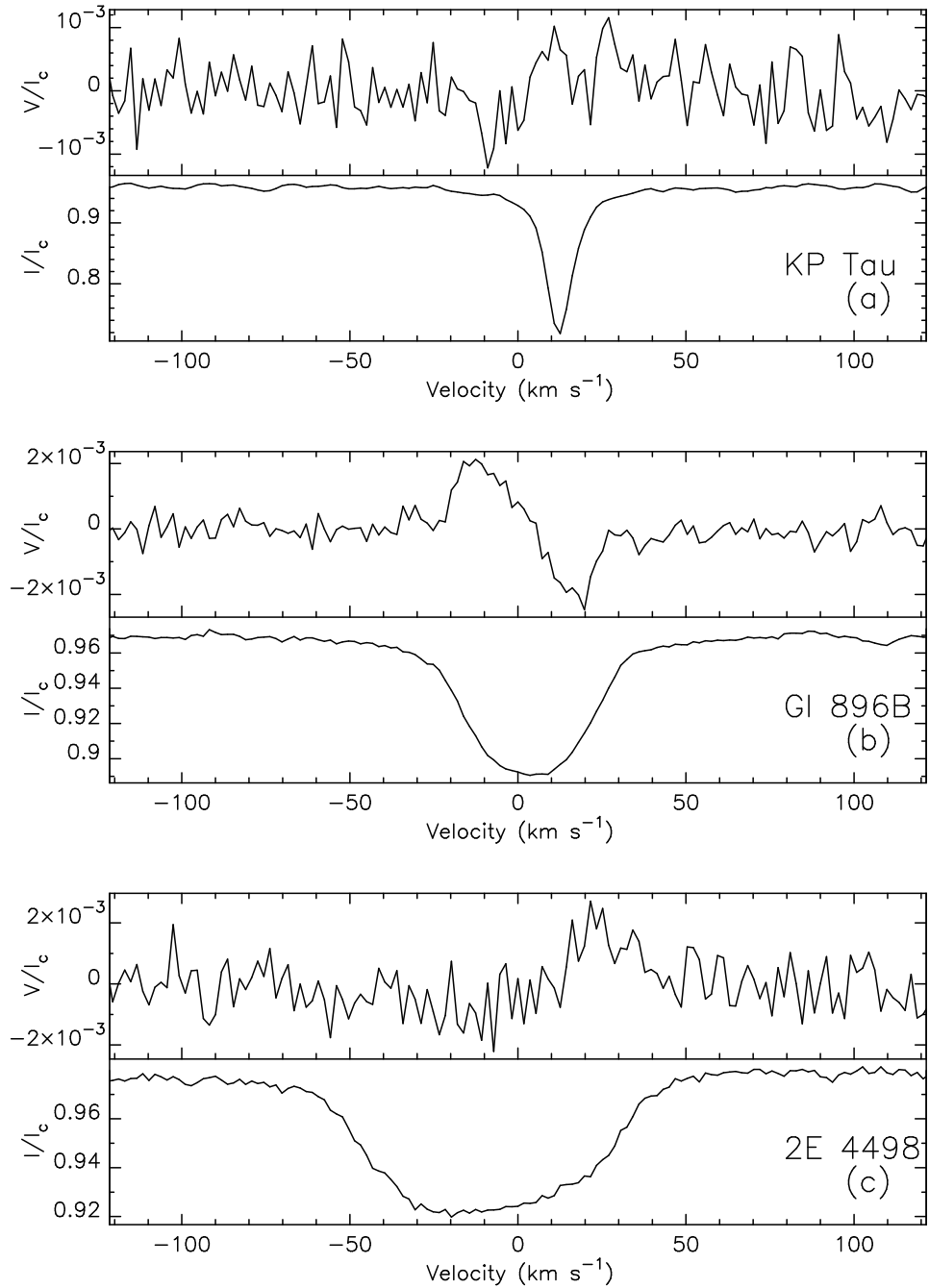}}
\caption{LSD Stokes $V$ and $I$ profiles 
of 3 M dwarfs: KP Tau (panel a), Gl 896B (b) and 2E 4498 (c). 
Strong circular polarization is seen in the three stars. 
\label{f1}}
\end{figure}

No spectral type estimate of this dwarf is available in the literature.
From our unpolarized (Stokes $I$) spectra obtained with ESPaDOnS (see \S~3.1), 
using a spectral index versus spectral type
relationships for M dwarfs (e.g., \citealt{martin99}),
we estimate its spectral type as M4.5$\pm$0.5, giving it an absolute magnitude
$M_{\rm I}\approx10.1$ (e.g., \citealt{leggett}). Comparison 
with the DENIS apparent magnitude $I=10.38$ \citep{epchtein}
then gives a distance of $\sim 11.4$~pc, assuming no reddening.  
The source shows coronal activity in X-ray emission (0.1-2.4 keV) with a measured flux
$f_{\rm X} = 4.52 \times 10^{-12}$~ergs~s$^{-1}$~cm$^{-2}$
and 20~cm radio emissions (see \citealt{brink} and references therein).
At the source distance, we derive log$L_{\rm X}=28.85$.
The radius of an M4.5 dwarf is about 0.3~R$_{\odot}$ \citep{chabrier97},
resulting in a maximum rotational period of 7.5~hours 
at $v$~sin$i$~=~55~km~s$^{-1}$ \citep{moch}.
We detected strong circular polarization in this dwarf (see Fig.~\ref{f1}). Since 
its rotation period is quite short, our detection
makes the source ($M_{\star}\sim0.2$~M$_{\odot}$) an excellent target for
future mapping of
the magnetic field on a fully convective star.

\section{SPECTROPOLARIMETRIC OBSERVATIONS AND RESULTS}
\subsection{Observations}
We observed the six M dwarfs described above with the Canada-France-Hawaii
Telescope's ESPaDOnS 
high-resolution spectrograph ($R = 65,000$; \citealt{donati03}), which 
provides a wavelength coverage of 370-1000~nm. 
The spectropolarimetric mode was used to provide
unpolarized (Stokes $I$) and circularly polarized (Stokes $V$)
spectra. Among the six targets, G~164-31 is a particularly interesting source since
it is an analog of V374~Peg (G 188-38), which is a rapidly rotating M4 dwarf
with $v$~sin$i \approx 37$~km~s$^{-1}$ showing a large-scale poloidal field \citep{donati06a}.
G~164-31 is therefore also an excellent target on which to study the field topology
in fully convective stars. 

Each observation consists of 4 exposures taken at different
polarimeter configurations and combined together to filter out 
spurious polarization signatures to first order \citep{donati97a}. 
Exposure times were computed using the exposure time
calculator for ESPaDOnS to obtain signal-to-noise
ratios above 100 at the wavelengths of interest.  This signal-to-noise
value is typical of what is needed to detect the circular
polarization in M dwarfs (e.g., \citealt{pb06}).
In the case of G~164-31, we monitored the star for a total of 20 hours 
from 2008 March 23 to 26 with 33 observations, aiming to cover at least one
full rotation period of the star. 
The full observing logs are given in Table~\ref{tab_obs}.

\begin{figure}
\vskip 1in
\hskip -0.25in
\centerline{\includegraphics[width=8.0in,angle=-90]{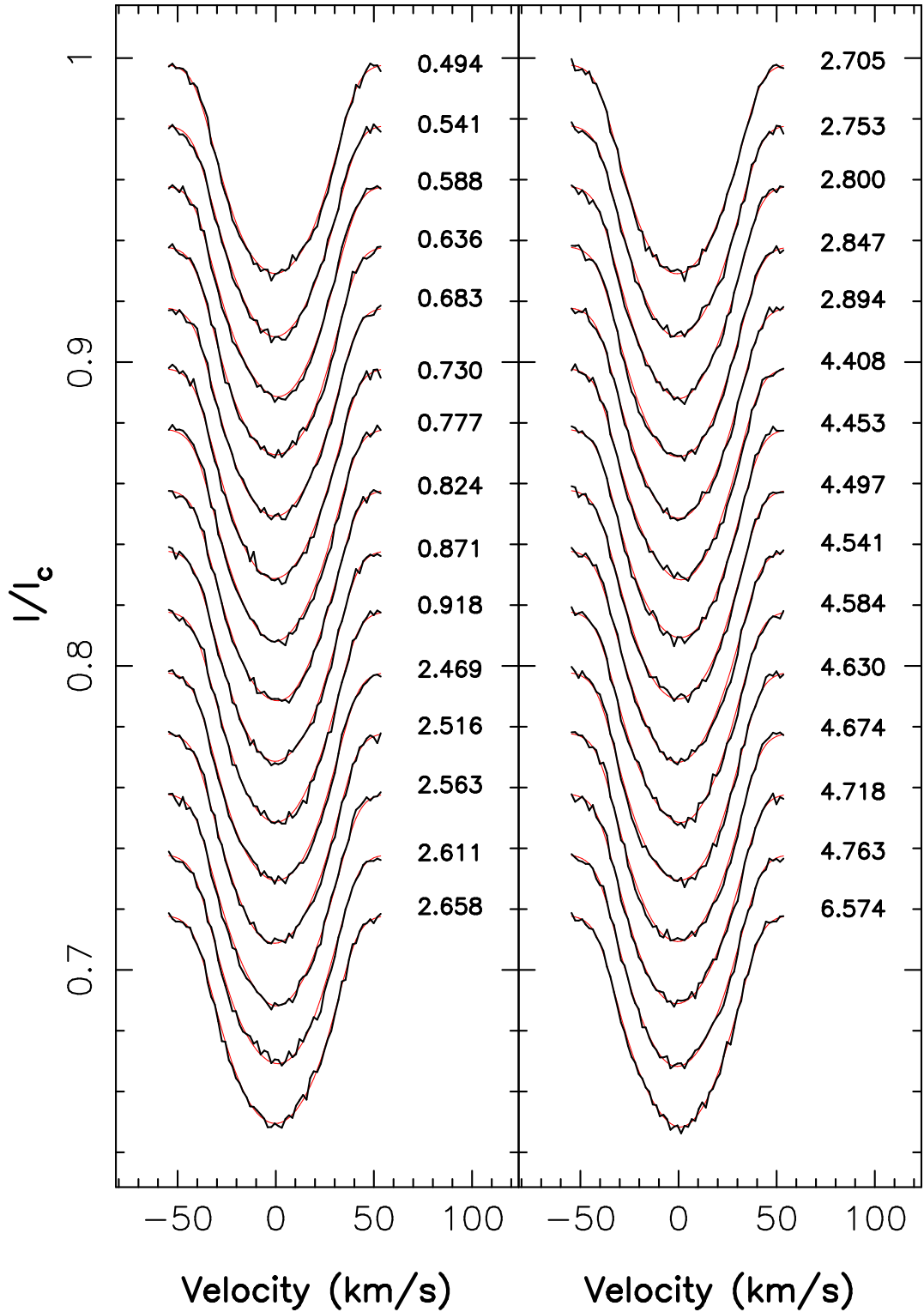}}
\caption{LSD Stokes $I$ profiles of G 164-31 (black line) along with the
maximum entropy fits (red line) to the data. The rotational phases 
are indicated to the right of each profile.
\label{f2}}
\end{figure}

\begin{figure}
\vskip 1in
\hskip -0.25in
\centerline{\includegraphics[width=8.0in,angle=-90]{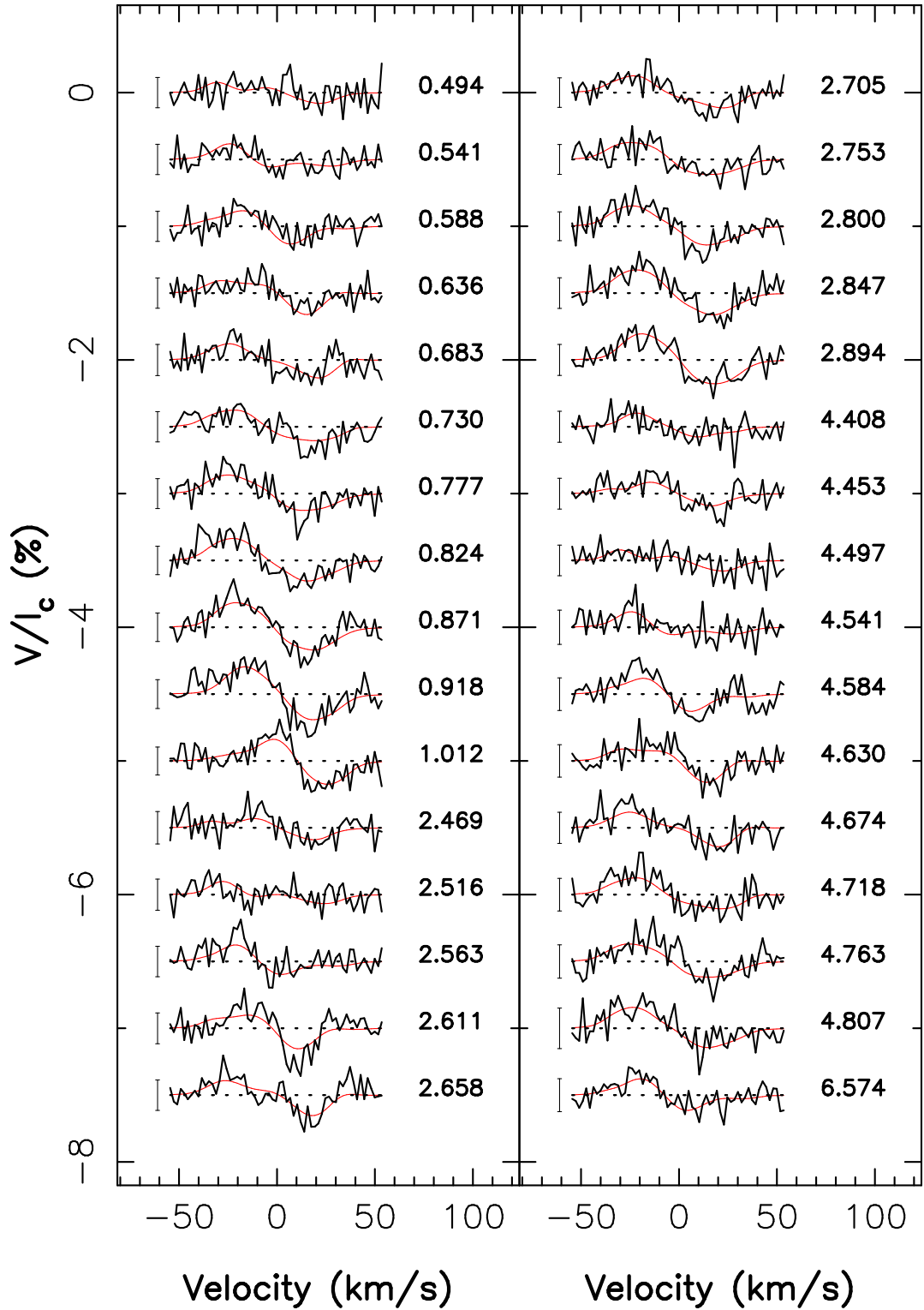}}
\caption{LSD Stokes $V$ profiles of G 164-31 (black line) along with
the maximum entropy fits (red line) to the data. The rotational phase ({\it right})
and the 3$\sigma$ error bar ({\it left}) of each profile are shown accordingly.
\label{f3}}
\end{figure}

Data reduction was performed using Libre-ESpRIT \citep{donati97a}
utilizing the principles of optimal extraction as given in \citet{horne}.
As the Zeeman signature is typically very weak, several methods
\citep{semel,gonzalez,donati97a} have been used in the past
to increase the signal-to-noise ratio of Zeeman signatures.
In this work, 
the Least-Squares Deconvolution (LSD, \citealt{donati97a}) 
multi-line analysis procedure
was applied to the spectra to extract the polarization signal from
a large number of photospheric atomic lines\footnote{
We used a line list \citep{kurucz} corresponding to M spectral 
types matching that of the six dwarfs.
Approximately 5000 intermediate to strong atomic spectral lines with an average Land\'e factor $g_{\rm eff}$
of $\sim1.2$ are used simultaneously to retrieve the average polarization information 
with typical noise levels of $\approx$0.08\% 
(relative to the unpolarized continuum level) per
1.8~km~s$^{-1}$ velocity bin and per individual polarization spectrum.  This
represents a 
multiplex gain in S/N of about 10 with respect to a single line analysis.} 
and compute a mean Zeeman signature corresponding
to an average photospheric profile centered at 700~nm.
Both mean 
Stokes $I$ and $V$ profiles are computed for all collected spectra.
Figure~\ref{f1} shows the LSD Stokes $I$ and $V$ profiles of KP~Tau, Gl~896B and 2E~4498
in which we detect Zeeman signatures. The profiles of G~164-31 are shown in
Figures~\ref{f2} and \ref{f3}.
Using the basic equations listed in \citet{wade} (see also \citealt{brown,donati97a}),
we compute the net longitudinal magnetic field strength, $B_z$, from our LSD Stokes $I$ and $V$ profiles.
In the case of G~164-31, its time series of Stokes~$I$ and Stokes~$V$ profiles are used in the
following section to reconstruct the field topology on this M4 dwarf.
\subsection{Results}
Table~\ref{tab_Bz} lists our measurements of $B_{\rm z}$ for all targets. 
We detected strong Zeeman signatures in KP~Tau, G164-31, Gl~896B and 2E~4498 but
we did not detect them in the two early-M dwarfs Gl~890 and LHS~473. 
The photospheric field strength is estimated from LSD Stokes $I$ and $V$ 
profiles,
and the chromospheric field is computed from the Stokes profiles 
of only the H$\alpha$ line.

For G~164-31, we find that the Zeeman signatures repeat
after 0.540$\pm$0.006 day 
(12.96$\pm$0.14 hours), which is in the expected range of
its rotation period. We thus conclude that the rotation period 
of this dwarf is 12.96~hours.
We note that
this value is slightly greater than the estimated maximum rotation period
of 12.1~hours for this star. The difference is possibly
due to G~164-31 having a larger radius than 
the expected value from theory.
Matching unpolarized (Stokes~$I$) spectra of G~164-31 with that of
slowly rotating references requires a
projected rotation velocity of $v$sin$i= 41\pm1$~km~s$^{-1}$. 
Our value is greater than one of \citet{fleming89}, 
$v$sin$i= 34\pm8.5$~km~s$^{-1}$, though both measurements are consistent
to within the uncertainties.  Dues to its smaller uncertainty, 
we adopt our $v$sin$i= 41$~km~s$^{-1}$ for G~164-31.
>From $P_{\rm rot}$ and $v$sin$i$, we derive $R$sin$i=0.44\pm0.02$~R$_{\odot}$.
Our $R$sin$i$ is $\sim30\%$ larger than the estimated stellar radius (see \S~2.1.4).  
It is unlikely that i exactly equals $90\degr$, and the results obtained in 
the imaging process decribed below are similar for 
$60\degr \leq i \leq 80\degr$.
We thus set $i=70\degr$, which results in the G~164-31 radius being
$\sim$38$\%$ larger than the typical value for M dwarfs of the 
same luminosity. This is likely due to the fact that G~164-31 is a young
object belonging to a moving cluster identified by \citet{orlov}.
We refer the reader to \S~5 for further discussion.

We monitored G~164-31 for 20 hours spread over three
successive nights, aiming for
coverage over a full rotational cycle. 
The true period (12.96~hours) is a bit longer than our expected maximum value
(12.1 hours). With this period, our observations cover only half of the 
rotational phase.
In other words, only half of the stellar surface was monitored. 
A few of the Stokes $I$ and $V$ LSD profiles are discarded due to the contamination of 
moonlight or flaring events. Figure~\ref{f2} and \ref{f3} show 
Stokes $I$ and $V$ LSD profiles of G~164-31.

\section{MODELING OF MAGNETIC FIELD TOPOLOGY OF G 164-31}
To reconstruct the map of cool spots and magnetic fields 
at the surface of G~164-31 from the set of observed LSD
Stokes $I$ and $V$, 
we use the \citet{donati01} magnetic mapping code 
which employes the Zeeman-Doppler imaging (ZDI) technique \citep{semel89}.
The typical longitude resolution achieved at the equator is $\approx15^{\circ}$
or $\approx0.04$ rotation cycle.
The basic idea of the code has been fully described in several papers
\citep{brown91,donati97b,donati01,donati06b,morin08}. 
In this paper, we briefly describe the principles of the modeling work
and refer the reader to the above papers for further details. 
\subsection{Mapping spots}
Cool spots on the surface of a rapidly rotating 
star produce distortions in the stellar spectral lines. If the star
is rotating fast enough (at least 20-30 km~s$^{-1}$ for late-type dwarfs, 
\citealt{vogt87}), 
the shape of the star's spectral line profiles
are dominated by rotational Doppler broadening. In this case, there is 
a strong correlation between the position of any distortion
within a line profile and the position of the corresponding spot on the
stellar surface. 
A spectrum of the rapid rotator is 
a one-dimensional image that is resolved in the direction
perpendicular to the stellar rotation axis and the light-of-sight. 
By observing the star at different rotation phases, 
a two-dimensional image of the spot distribution can be reconstructed
using the so-called Doppler imaging (DI) technique 
(see \citealt{vogt87} and references therein).
In the imaging process, the stellar surface is divided into a grid
of 1000 elementary cells. 
Using the spot-occupancy model of \citet{cameron92},
the local line profile at each grid point
of the surface is described as a linear combination of two 
reference profiles, one representing the quiet photosphere and one for cool spots.
Both reference profiles are assumed to be equal and only differ
by their relative continuum levels. For the assumed reference profile, we can use
either the LSD profile of the very slowly rotating inactive M4 dwarf Gl~402
or a simple Gaussian profile with similar full width at half-maximum and equivalent
width. Both options yield very similar results, indicating that the exact shape
of the assumed local profiles has very minor impact on the reconstructed images, as long as the rotational velocity of the star is much larger than the local profile.
Each element on the surface is quantified by the local fraction occupied by spots.
The spot occupancy ranges from 0 (no spots) to 1 (complete spot
coverage). The maximum entropy image reconstruction technique \citep{ski,vogt}
is used to find the image having the smoothest variation (or in some
sense the simplest image), giving
the least contrast between spots and the quiet photosphere.

\subsection{Mapping magnetic fields}
The field is described as the sum of a poloidal and a toroidal component,
both expressed as spherical harmonics expansions (\citealt{jardine99}).
For a given set of the complex coefficients 
of the spherical harmonic expansions, one can
produce a corresponding topology (or a Stokes $V$ data set) at the stellar
surface.
The code uses the maximum entropy image reconstruction technique,
in which entropy (i.e., quantifying the amount of reconstructed information)
is calculated from the coefficients of the spherical harmonics expansions.
The imaging process starts from a
null magnetic field and iteratively adjusts the spherical harmonic 
coefficients in order to match the synthetic Stokes $V$ profile to the 
observed LSD one.  The code uses
a multidirectional search in the image space until 
the required maximum entropy image is obtained.
This corresponds to an optimal field topology that 
reproduces the data at a given $\chi^{2}$ level,
i.e., usually down to noise level. 
Since the inversion problem is partially
ill posed, we use the entropy function to select the magnetic field map
with lowest information content among all those reproducing the
data equally well. 
 
To calculate the synthetic Stokes $V$ profiles for a given field topology, 
the surface was again divided into a grid of 1000 elementary surface cells.
In each cell, the three components
(radial, azimuthal and meridional in spherical coordinates) of the vector field 
are estimated  directly from the spherical harmonic expansion.
The code uses the analytic solutions \citep{lando} 
of Unno-Rachkovsky's radiative transfer equations 
to calculate the contribution to the Stokes $V$ profiles of 
all visible cells
at each observed rotation phase. 
The free parameters in the Unno-Rachkovsky equations are obtained
by fitting the LSD Stokes $I$ profile of a very slowly rotating
and weakly active star with a similar spectral type (e.g., Gl 402).
To obtain the best fit (in both amplitude and width)
between the synthetic and observed Stokes $V$ profiles,
the code introduces a filling factor $f_{\rm c}$, which represents
the fractional amount of circular flux being constant over the whole
stellar surface, and speculates that large-scale fields in 
M dwarfs can be structured on a small scale \citep{donati08}.
The circularly polarized flux
from each cell is thus $f_{\rm c}V_{\rm loc}$, where $V_{\rm loc}$
is the Stokes $V$ profile derived from the analytic solutions
of the Unno-Rachkovsky equations.

\begin{figure}
\vskip 1in
\hskip -0.25in
\centerline{\includegraphics[width=3.5in,angle=0]{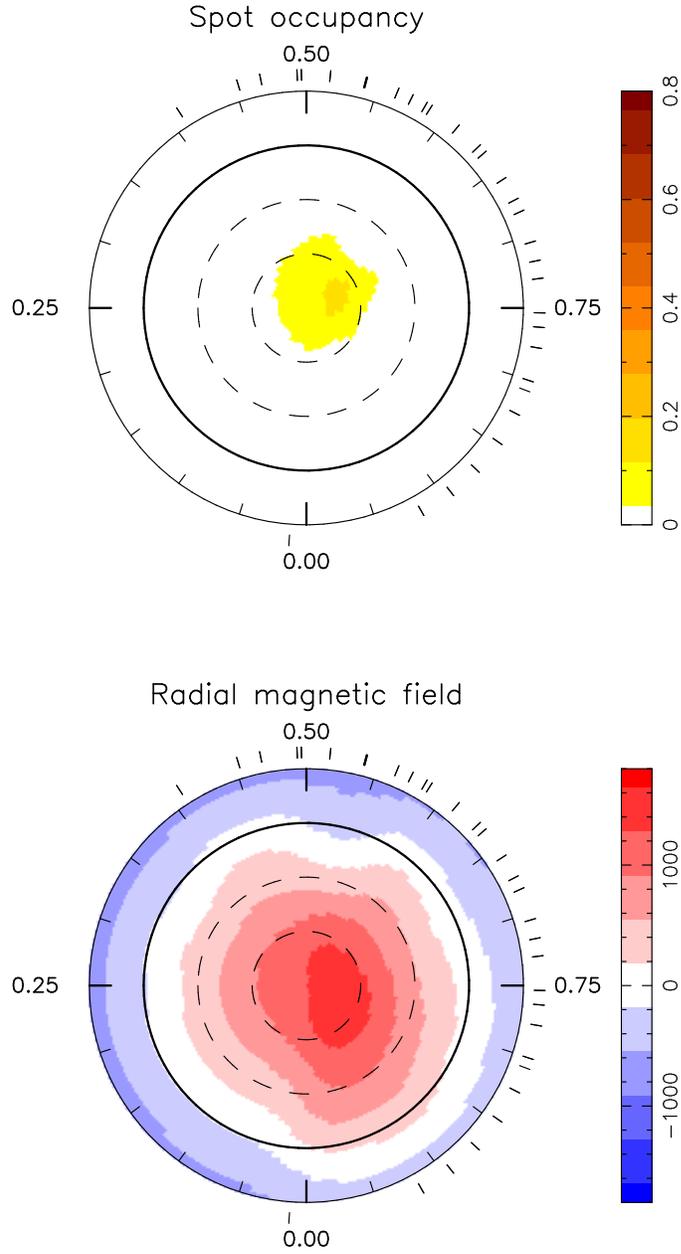}}
\caption{Magnetic field topologies of G 164-31, reconstructed from the series of LSD Stokes $V$ profiles.
The radial field ({\it bottom}) dominates in the dwarf
(intensity scale in Gauss), 
the azimuthal and meridional component are negligible.
The spottedness ({\it top}) at the stellar surface is also reconstructed, the color scale 
represents spot occupancy for the dwarf (1$=$complete spot coverage).
The dwarf is shown in flattened polar projection extending down to latitudes of $-$30$\degr$.
The equator is described as a bold circle. Radial ticks outside the plots
indicate the phases at which the dwarf was observed, a half of the dwarf surface had properly been monitored.
\label{f4}}
\end{figure}

Differential rotation is also implemented
in the calculation process of the synthetic Stokes V profiles
for yielding the best fits to the observation.
For this purpose, 
we assume the rotation rate varies with latitude, $\theta$,
following a solar-like differential rotation law 
as $\Omega(\theta)=\Omega_{\rm eq}-d\Omega$~sin$^{2}\theta$, where
$\Omega_{\rm eq}$ is the angular rotation rate at the equator, and 
$d\Omega$ is the difference in angular rotation rate between the equator
and the pole. Differential rotation is detected when 
$\chi^{2}$ of the fit to the data shows a well defined minimum
in the range of $\Omega_{\rm eq}$ and $d\Omega$ values.
In the case of G~164-31, we did not obtain
a clear minimum in the explored $\Omega_{\rm eq}$ $-$ $d\Omega$
range. This indicates that our data
are not suitable for measuring differential rotation
given the fairly simple (and low amplitude) rotational modulation of 
the Stokes~$V$ profiles we observe.
This is mostly due to the fact 
that the field distribution lacks well defined features at different 
latitudes from which differential rotation can be estimated.  
Based on the previous observations of analogs
of G~164-31 \citep{donati06a,morin08},
we therefore assumed that this star rotates as a rigid body  while
performing the field mapping.
The obtained map of G~164-31 
is shown in Figure~\ref{f4}.  %
\subsection{Results}
Figure~\ref{f4} presents our maps of the spot occupancy and
the magnetic field at the stellar surface
of G~164-31. For the half stellar surface that was not observed, 
the field topology has been derived using a spherical harmonic expansion based on the collected Stokes $V$ 
data set. 
Our brightness map shows a low contrast spot at the pole of the star.
The magnetic field morphology 
on G~164-31 shows that
the radial field component is dominant, 
which appears similar to what is seen in V374~Peg. 
The azimuthal and meridional components are negligible. 

The average large-scale flux is 680~G and the poloidal field
dominates with 99$\%$ of the energy content. Most (95$\%$)
of this poloidal field is stored in the mode
corresponding to a dipole aligned with the rotation axis
(spherical harmonics degree $l=1$ and order $m=0$).
One should note that 
due to the high $v$sin$i$ of G~164-31, for any value of the filling factor
$0 < f_{\rm c} < 1$, we find similar results. Hence, we arbitrarily 
set $f_{\rm c} = 1$. 
We also examine the binarity of G~164-31. 
Based on the LSD Stokes $I$ profiles, we find an average radial velocity value 
of $-6.7\pm0.2$~km~s$^{-1}$.  
No significant variation in radial velocity is found. Therefore,
our current data does not reveal any hint of an
additional close-in component around G~164-31.

\section{DISCUSSION}
In this section, we first briefly discuss the magnetic field 
properties of the stars whose fields are not mapped. These stars are
Gl~890, LHS~473, KP~Tau, Gl~896B and 2E~4498. We then focus 
the discussion on the field topology in the photosphere
and the chromosphere of G~164-31.
\subsection{Individual stars: Gl 890, LHS~473, KP Tau, Gl 896B and 2E 4498}
We did not detect Zeeman signatures in Gl~890 and
LHS 473. For the latter case, the non-detection in both the photosphere
and the chromosphere indicates that
LHS~473 is inactive, likely due to its low $v$sin$i$, and this result is 
consistent with previous observations showing its low coronal activity level. 
However in the fastly rotating case of Gl~890, 
the non-detection suggests that the field topology
on this star may be different from what has been observed 
in more slowly-rotating early-M dwarfs \citep{donati08},
whose fields are large-scale and
dominantly toroidal producing Stokes~V signatures detectable at any time.
Since the detections of X-ray and radio emission indicate that
Gl~890 is magnetically active,
phase-resolved spectropolarimetric observations of this star
are therefore needed to clarify its magnetic field topology.

For the three remaining M dwarfs, all these
fast rotators show Zeeman signatures, and they are
particularly strong in Gl~896B and 2E~4498 (Fig.~\ref{f1}).  This result
indicates the presence of strong, large-scale fields on these stars. 
Since KP~Tau (0.31~M$_{\odot}$) and 2E~4498 (0.21~M$_{\odot}$) are fully convective, 
they are good targets for
studying the field morphology by carrying out spectropolarimetric 
and simultaneous multiple-wavelength observations (e.g., \citealt{berger09}).
We note that Gl~896Bab is an unresolved binary. 
Therefore, its Zeeman signature is contributed by two components if both are
magnetically active.

\subsection{G 164-31}
The G~164-31 magnetic field topology, which is mostly poloidal
with most of the magnetic energy concentrated within
the lowest order axisymmetric modes, is very similar to
what has been observed in V374~Peg \citep{donati06a}
and other mid-M dwarfs \citep{morin08}.
The average large-scale magnetic flux on G~164-31 is 0.68~kG, which
is significantly smaller than the previous measurements
of few kiloGauss magnetic fluxes in active mid-M dwarfs 
using synthetic spectrum fitting techniques
(e.g., \citealt{jk96}).  
This is possibly due to the ZDI technique being sensitive
to only large-scale and simple structures, while Zeeman signatures from 
magnetic regions with complicated or small-scale topology may cancel
each other in circularly polarized spectra whereas these signatures 
add up in unpolarized spectra. To explore this 
``missing" magnetic flux, we used the synthetic spectrum fitting technique
described in \citet{jk96} to measure the mean magnetic flux on the
surface of G~164-31.  Briefly, spectra in the wavelength interval around 
the Zeeman-sensitive Fe~{\small I} line at 8468.40~\AA~ are analyzed. 
In stars as cool as G~164-31, this line is actually blended with 
a Ti~{\small I} line at 8468.47~\AA~ which is similar in strength and
also Zeeman sensitive.  The spectrum synthesis models both lines plus
numerous TiO lines in this wavelength region.  Because of the ubiquitous
TiO, spectral changes due to magnetic fields are seen more
clearly in the ratio of active to inactive line profiles.  Therefore,
spectra of G~164-31 were divided by inactive references.
In our case, we used two inactive reference stars 
GJ~725B (or LHS~59, M3.5V) and GJ~876 (or LHS~530, M4V) which are
meant to be identical in all respects ($T_{eff}$, log$g$, [M/H], etc.)
to G~164-31 except for the field.  Line profiles were synthesized with 
and without magnetic fields (see \citealt{jk96} and references therein) 
and the ratio of active to inactive profiles was fit by adjusting the
magnetic field strength, $B$, and its filling factor, $f$, in the
model spectrum of G~164-31.  The line profiles are generated assuming
a uniform radial field everywhere with a uniform filling factor.

The best fits were obtained with $B f=3.1$~kG and $B f=3.3$~kG for GJ~725B 
and GJ~876.  The fit to the observations using GJ~876 as the inactive
reference star is shown in Figure 5.  Due to the high $v$sin$i$ of G~164-31, 
we can not measure $B$ and $f$ separately.  The model fit in the line
core is not as good as we would like, and is only slightly improved using
GJ~725B as the reference star.  We note though that if the radius of G~164-31
is indeed larger than expected due to youth (see below), its gravity is
likely somewhat lower than that of either inactive comparison star.  Our
spectrum synthesis indicates that the 8468~\AA~ feature should actually
weaken somewhat at lower gravity, which would result in the discrepancy in
the core getting worse by $\sim 0.005$ in the ratio.  A possible way to
correct for this would be to fit a distribution of magnetic field strengths
on the stellar surface as has been done for M dwarfs (Johns--Krull \&
Valenti 2000) and on T Tauri stars (e.g. Johns--Krull et al. 1999).  The
current fit (Fig. 5) attempts to use a single field strength to fit both
the wings and the core, whereas a distribution of magnetic fields gives
more flexibility in fitting the entire profile and can result in bigger
changes in the line equivalent width than produced when using only a
single field value.  We therefore adopt $Bf =3.2\pm0.4$~kG for G~164-31
as the best estimate from our single field fitting described above, noting
that this may be a slight underestimate.  This value for $Bf$ is larger 
than the one measured from the ZDI technique by a factor of 4.7.
Our result is consistent with the previous measurements from the literature 
(e.g., EV~Lac, \citealt{jk96,morin08}), suggesting that in active mid-M 
dwarfs a significant part of the magnetic energy is stored in small-scale
structures.  As found by \citet{rn07}, it appears that a large majority of
the magnetic energy is stored in the small scale field of G~164-31.
Apparently, a successful dynamo model should be able to
produce both large-scale poloidal fields
and small-scale features in rapidly rotating mid-M dwarfs.

To study chromospheric activity, we computed the H$\alpha$
emission equivalent widths for each exposure. 
Our H$\alpha$ emission light curve is shown in Fig.~\ref{f6}.
One should keep in mind that since each observation to obtain a Stokes $V$ profile
consists of 4 exposures, measuring H$\alpha$
emission at individual exposures therefore
increases time resolution of the light curve.
Figure~\ref{f6} also indicates that the H$\alpha$ equivalent width variations
are likely sinusoidal and they are modulated
by the stellar rotation with $P_{\rm rot}=0.54$~day determined from
the Stokes~$V$ profiles. This effect has also been observed
in late-M and brown dwarfs (see \citealt{berger09} and references therein).
Flaring events were seen strongly in the last sequences 
of the first observing night and moderately in the middle sequences 
of the second night. These events were also observed in the H$\beta$ emission profiles.
While the likely rotational modulation and nonzero minimum of the H$\alpha$ light curve
imply either a large-scale (poloidal or toroidal) 
chromospheric field or a localized concentration
of small-scale magnetic activity, it is reasonable to assume
that the chromospheric field in G~164-31 has both a large scale
poloidal component and small-scale structures as observed in the photospheric field. 
This field configuration probably dominates in both
the photosphere and the chromosphere of the star. 
We note that
simultaneous observations at multiple wavelengths
of the source will place more contraints on the chromospheric
and coronal field morphology (e.g., \citealt{berger09}).

It is very interesting to note that the radius of G~164-31 is at least
30$\%$ larger than that estimated from standard models for M dwarfs at the same 
luminosity (see \S~2.1.4). There are two possible interpretations of this discrepancy.
First, the magnetic field effects of reduced convective efficiency 
due to fast rotation and large field strengths, and/or to spot coverage 
as proposed by \citet{chabrier07} might yield a cooler effective temperature $T_{\rm eff}$
and thus a larger radius in low-mass stars, 
keeping their luminosity unchanged
($L \propto T^{4}_{\rm eff}R^{2}$).
Our spot mapping (Fig.~\ref{f4}) shows the fraction of the stellar surface covered
by the low contrast spot is relatively small in G~164-31. However, one 
should note that the imaging technique is only sensitive to 
spots or spot groups with sizes comparable to our resolution element. 
We therefore can not rule out the possibility of unresolved small spots spread 
everywhere on the stellar surface. Hence,
we can not conclude whether the effect due to spot coverage 
is significant in this star or not. 
%The latter one (reduced convective efficiency)
%is expected to yield a $\sim7\%$ larger radius.
Observational results (see \citealt{ribas,morales} and references therein) suggest that 
these effects might yield a $\sim10-15\%$ larger radius in G~164-31,
therefore the discrepancy in radius of over $30\%$ can not be explained
by only those effects.
Second, we explore the possibility that
G~164-31 is a young M4 dwarf, its radius thus larger than one
computed for old M dwarfs. We note that the $K$-band absolute
magnitude of G~164-31, $M_{\rm K}=6.73$, is 0.67~mag brighter than
the average value for M4 dwarfs\footnote{We selected 9 single M4 dwarfs within 10~pc
listed in \cite{del98} and computed their $K$-band absolute magnitude
from trigonometric parallaxes and 2MASS $K$-band magnitudes available in the VIZIER database. 
This work yielded an average value $M_{\rm K}=7.4$.}, supporting the young nature of
the star. From a literature search, 
the binary system G~164-31 (Gl 490B)$+$G~164-32 (Gl~490A) 
indeed belongs to a moving cluster previously identified 
by \citet{orlov} (for a review on young nearby stars, see \citealt{zuckerman}). 
One should note that 
the primary component G~164-32 also exhibits high coronal X-ray emission with
log($L_{\rm X}$/$L_{\rm bol}$)$=-3.23$ \citep{fleming89} 
and chromospheric H$\alpha$ emission \citep{stauffer86}, 
indicating activity typical in young early-M dwarfs.
We measured an upper limit for the Li~$\lambda6708$ equivalent width of
17~m\AA~for G~164-31. This gives the star an age older than 10~Myr
since early-M dwarfs ($\geq$M4) are expected to
completely deplete their Lithium within $\sim$10~Myr.
We suggest G~164-31 is an analog of the M4 dwarf HIP~112312A at 23.6~pc \citep{esa}, 
which is a member of the $\sim$12 Myr old $\beta$ Pictoris moving group \citep{song}.
To estimate the mass and age of G~164-31, 
we used the \citet{chabrier07,b98} theoretical models with constraints of 
$M_{\rm K}=6.73$, $I-J=1.61$, and the deduced radius $R_{\star}=0.47$~R$_{\odot}$ of G~164-31,
and we also considered the magnetic field effects that might yield about a 10-15$\%$ larger radius. 
We then derived $M_{\star} \sim 0.15$~M$_{\odot}$ at an age of about 25-30~Myr,
making G~164-31 the least massive M dwarf whose magnetic field has been
mapped to date.
More observations are needed to precisely determine the fundamental parameters
of this moving cluster.

\section{SUMMARY}
Based on our circularly polarized and unpolarized spectroscopic observations,
using both tomographic imaging and synthetic spectrum fitting,
we reveal the mainly axisymmetric large-scale poloidal magnetic field  
and the small-scale field structures storing a significant portion of the magnetic
energy in the photosphere of the M4 dwarf G~164-31. The modulation
of the H$\alpha$ emission light curve suggests that the field in the chromosphere
is stable and possibly poloidal like that in the photosphere.
Our detection of circular polarization in the single and rapidingly rotating 
M dwarfs KP~Tau (M3V), 2E~4498 (M4.5V) make them
good targets for mapping of the field morphology in mid-M dwarfs 
in the future.

\acknowledgments
N. P.-B. has been aided in this work by a 
Henri Chretien International Research Grant administered by
the American Astronomical Society.
E. M. acknowledges support from the Spanish Ministry of Science through the   
project AYA2007-67458. The authors extend special thanks to those of   
Hawaiian ancestry on whose sacred mountain we are privileged to be guests.
Access to the CFHT was made possible by the Ministry of Education
and the National Science Council of Taiwan as part of the
Cosmology and Particle Astrophysics (CosPA) initiative.

\clearpage

\begin{deluxetable}{lllllllrrl}
\tablewidth{0pt}
\tabletypesize{\footnotesize}
\tablecaption{Physical parameters of the 6 M dwarfs
  \label{tab_Bz}}
\tablehead { 
\colhead {Star  }          &  
\colhead {SpT\tablenotemark{a}} & 
\colhead {$M_{\rm K}$\tablenotemark{b}} & 
\colhead {log$L_{\rm bol}$\tablenotemark{c}} & 
\colhead {$v$sin$i$\tablenotemark{d}} &
\colhead {log($R_{\rm X}$)\tablenotemark{e}} & 
\colhead {$M_{\star}$\tablenotemark{f}} &
\colhead {$B_{\rm z}$} & \colhead {H$\alpha$ EW} & \colhead {$B_{\rm z}$(H$\alpha$)}     \\
\colhead {}  & 
\colhead {} &
\colhead {} &
\colhead {(ergs s$^{-1}$)} & 
\colhead {(km~s$^{-1}$)} &
\colhead {}&
\colhead {(M$_{\odot}$)}   & 
\colhead {(G)}& \colhead {(\AA)} & \colhead {(G)} 
}
\startdata
Gl 890 & M0    & 5.41 & 32.26 & 70     &$-$3.26 &  0.57 & $<$18~\,~\,~\,&    0.64$\pm$0.01 & ~\,$<$22 \\
LHS 473& M2.5  & 5.83 & 32.09 & $<$2.6 &$-$4.98 &  0.48 &    16$\pm$7~\,& $-$0.29$\pm$0.02 & ~\,$<$30 \\
KP Tau & M3    & 6.82 & 31.66 & 32     &$-$3.53 &  0.31 & $-$104$\pm$32 &    0.73$\pm$0.05 & $-$580$\pm$220 \\
       &       &      &       &        &        &       &$-$92$\pm$32   &    0.73$\pm$0.05 & ~~\,210$\pm$110 \\
G 164-31&M4    & 6.73 & 31.66 & 41     &$-$2.61 &  0.33 &  680\tablenotemark{g}~\,~\,&    3.15$\pm$0.01\tablenotemark{g} & ~~\,530$\pm$60\tablenotemark{g} \\
Gl 896B& M4.5  & 7.28 & 31.45 & 24.2   &$-$3.26 &  0.25 & 296$\pm$40    &    3.5$\pm$0.3~\,   & $-$150$\pm$60 \\
       &       &      &       &        &        &       & 294$\pm$37    &    3.6$\pm$0.3~\,   & ~~\,110$\pm$60 \\
       &       &      &       &        &        &       &279$\pm$38     &    4.8$\pm$0.5~\,   & ~~\,160$\pm$15 \\
2E 4498& M4.5  & 7.60 & 31.31 & 55     &$-$2.46 &  0.21 &$-$440$\pm$84  &    4.7$\pm$0.5~\,   & $-$270$\pm$100 \\
\enddata
\tablenotetext{a}{Spectral type from the literature.}

\tablenotetext{b}{Absolute $K$-band magnitudes computed from trigonometric parallaxes (Gl 890, LHS 473, G 164-31, Gl 896B) or spectroscopic distances (KP Tau, 2E 4498) and 2MASS $K$-band magnitudes available in the VIZIER database.}

\tablenotetext{c}{Bolometric luminosity estimated using the bolometric correction $BC_{\rm K}$ versus
$K$-band magnitude relationship in \citet{tinney}.}

\tablenotetext{d}{Rotational velocity, see text (\S~2.1) for references, except G 164-31 (this paper).}

\tablenotetext{e}{$R_{\rm X}=L_{\rm X}$/$L_{\rm bol}$.}

\tablenotetext{f}{Stellar masses estimated using the $M_{\rm K}$ versus mass empirical 
relationship in \citet{delfosse00}.}

\tablenotetext{g}{For G 164-31: the value listed for an average magnetic flux
$Bf_{\rm c}$ computed using the ZDI technique (see \S~4.3), assuming $f_{\rm c}=1$. Its H$\alpha$ equivalent width
and chromospheric field strength $B_{\rm z}$ are estimated at cycle $E=0.9181$ where the Zeeman signature
appears very strong (see Fig.~\ref{f3}).}

\end{deluxetable}
\clearpage

\begin{deluxetable}{lrrcccc}
\tablewidth{0pt}
\tablecaption{Observing logs for the 6 M dwarfs
  \label{tab_obs}}
\tablehead { 
\colhead {Target}          & \colhead {Date} & \colhead {HJD\tablenotemark{a}} & \colhead {UT} &
\colhead {Exposure}        & \colhead {Peak of}           & \colhead {Cycle\tablenotemark{b}}   \\
\colhead {}                & \colhead {} & \colhead {(2453000+)} &\colhead {(h:m:s)} & \colhead {time (s)} &
\colhead {S/N} & \colhead {}  
}
\startdata
Gl 890    & 2007 Sep 29 & 1372.91386 & 09:49:13 & 4$\times$200 & 166 &  ...    \\
LHS 473   & 2005 Sep 18 &  631.74709 & 05:52:12 & 4$\times$300 & 511 &  ...    \\
KP Tau    & 2007 Sep 29 & 1373.05187 & 13:08:24 & 4$\times$500 & 153 &  ...    \\
          &          29 & 1373.07920 & 13:47:45 & 4$\times$500 & 153 &  ...    \\
Gl 896B   & 2007 Sep 29 & 1372.93533 & 10:18:07 & 4$\times$300 & 182 &  ...    \\
          &          29 & 1372.95603 & 10:47:55 & 4$\times$400 & 209 &  ...    \\
          &          29 & 1373.02336 & 12:24:53 & 4$\times$400 & 195 &  ...    \\
2E 4498   & 2007 Sep 29 & 1372.89021 & 09:15:32 & 4$\times$500 & 139 &  ...    \\  
G 164-31  & 2008 Mar 24 & 1549.80662 & 07:16:47 & 4$\times$500 & 140 & 0.4937  \\
          &          24 & 1549.83227 & 07:53:43 & 4$\times$500 & 144 & 0.5412  \\
          &          24 & 1549.85770 & 08:30:19 & 4$\times$500 & 145 & 0.5883  \\
          &          24 & 1549.88332 & 09:07:13 & 4$\times$500 & 143 & 0.6358  \\
          &          24 & 1549.90865 & 09:43:42 & 4$\times$500 & 140 & 0.6827  \\
          &          24 & 1549.93419 & 10:20:28 & 4$\times$500 & 141 & 0.7300  \\
          &          24 & 1549.95967 & 10:57:10 & 4$\times$500 & 144 & 0.7772  \\
          &          24 & 1549.98510 & 11:33:47 & 4$\times$500 & 145 & 0.8243  \\
          &          24 & 1550.01040 & 12:10:13 & 4$\times$500 & 145 & 0.8711  \\
          &          24 & 1550.03576 & 12:46:44 & 4$\times$500 & 144 & 0.9181  \\
          &          24 & 1550.06109 & 13:23:13 & 4$\times$500 & 149 & 0.9650  \\
          &          24 & 1550.08637 & 13:59:36 & 4$\times$500 & 150 & 1.0118  \\
          &          25 & 1550.87304 & 08:52:24 & 4$\times$500 & 132 & 2.4686  \\
          &          25 & 1550.89882 & 09:29:31 & 4$\times$500 & 138 & 2.5163  \\
          &          25 & 1550.92424 & 10:06:08 & 4$\times$500 & 144 & 2.5634  \\
          &          25 & 1550.94977 & 10:42:53 & 4$\times$500 & 137 & 2.6107  \\
          &          25 & 1550.97519 & 11:19:29 & 4$\times$500 & 140 & 2.6578  \\
          &          25 & 1551.00055 & 11:56:01 & 4$\times$500 & 149 & 2.7047  \\ 
          &          25 & 1551.02679 & 12:33:48 & 4$\times$500 & 144 & 2.7533  \\
          &          25 & 1551.05210 & 13:10:15 & 4$\times$500 & 145 & 2.8002  \\
          &          25 & 1551.07741 & 13:46:41 & 4$\times$500 & 135 & 2.8471  \\
          &          25 & 1551.10286 & 14:23:20 & 4$\times$500 & 145 & 2.8942  \\
          &          26 & 1551.92059 & 10:00:51 & 4$\times$464 & 138 & 4.4085  \\
          &          26 & 1551.94449 & 10:35:15 & 4$\times$464 & 139 & 4.4528  \\
          &          26 & 1551.96815 & 11:09:20 & 4$\times$464 & 144 & 4.4966  \\
          &          26 & 1551.99187 & 11:43:30 & 4$\times$464 & 127 & 4.5405  \\
          &          26 & 1552.01563 & 12:17:42 & 4$\times$464 & 132 & 4.5845  \\
          &          26 & 1552.04024 & 12:53:09 & 4$\times$464 & 134 & 4.6301  \\
          &          26 & 1552.06392 & 13:27:14 & 4$\times$464 & 141 & 4.6739  \\
          &          26 & 1552.08759 & 14:01:20 & 4$\times$464 & 134 & 4.7178  \\
          &          26 & 1552.11210 & 14:36:37 & 4$\times$464 & 132 & 4.7631  \\
          &          26 & 1552.13587 & 15:10:51 & 4$\times$464 & 114 & 4.8072  \\
          &          27 & 1553.08983 & 14:04:32 & 4$\times$464 & 135 & 6.5738  \\
\enddata
%\tablecomments{}
\tablenotetext{a}{Heliocentric Julian date (UTC).}

\tablenotetext{b}{Rotational cycles $E$ are computed
using ephemeris HJD~$=2454549.54+0.54 E$, only available for G 164-31.}

\end{deluxetable}

\clearpage

\clearpage

\clearpage
\begin{figure}
\vskip 1in
\hskip -0.25in
\centerline{\includegraphics[width=4.5in,angle=90]{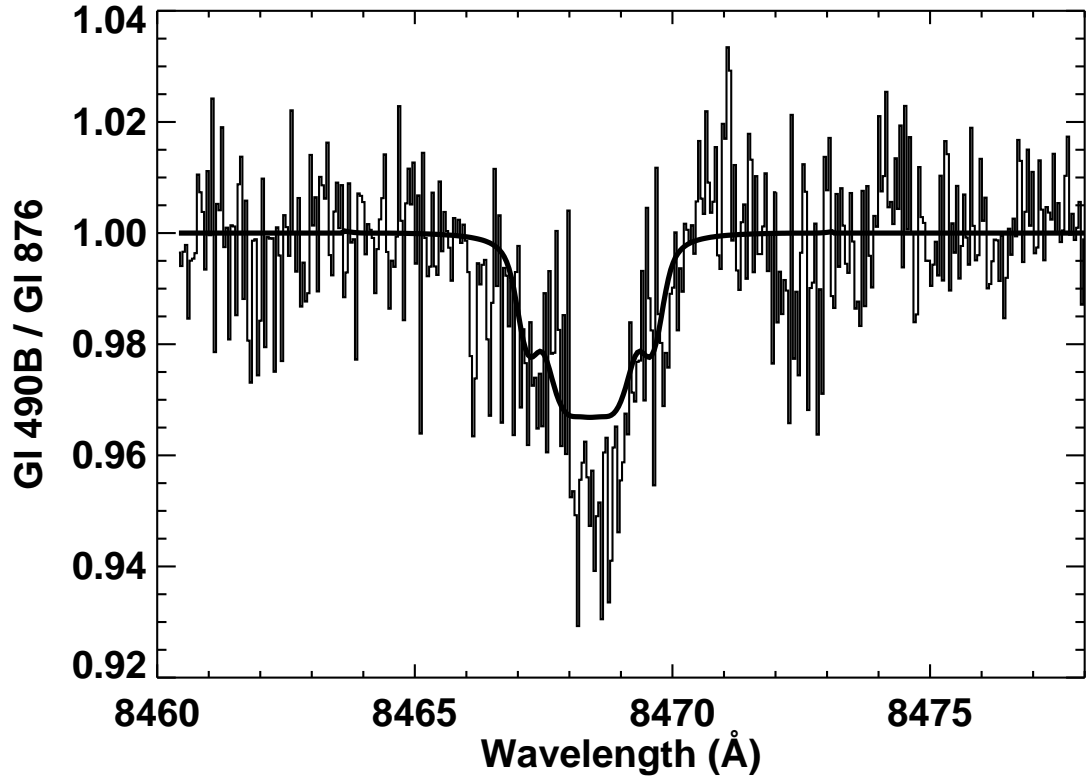}}
\caption{Best-fit modeling ratio of active to inactive profiles (thick line).
The observed line profile ratio constructed by dividing the spectrum of G~164-31 (Gl 490B)
in the wavelength region around the magnetically sensitive
Fe~{\small I} line at 8468.4~\AA~by that
of the inactive M4 dwarf Gl 876 is shown (thin line).
\label{f5}}
\end{figure}

\clearpage
\begin{figure}
\vskip 1in
\hskip -0.25in
\centerline{\includegraphics[width=3.8in,angle=-90]{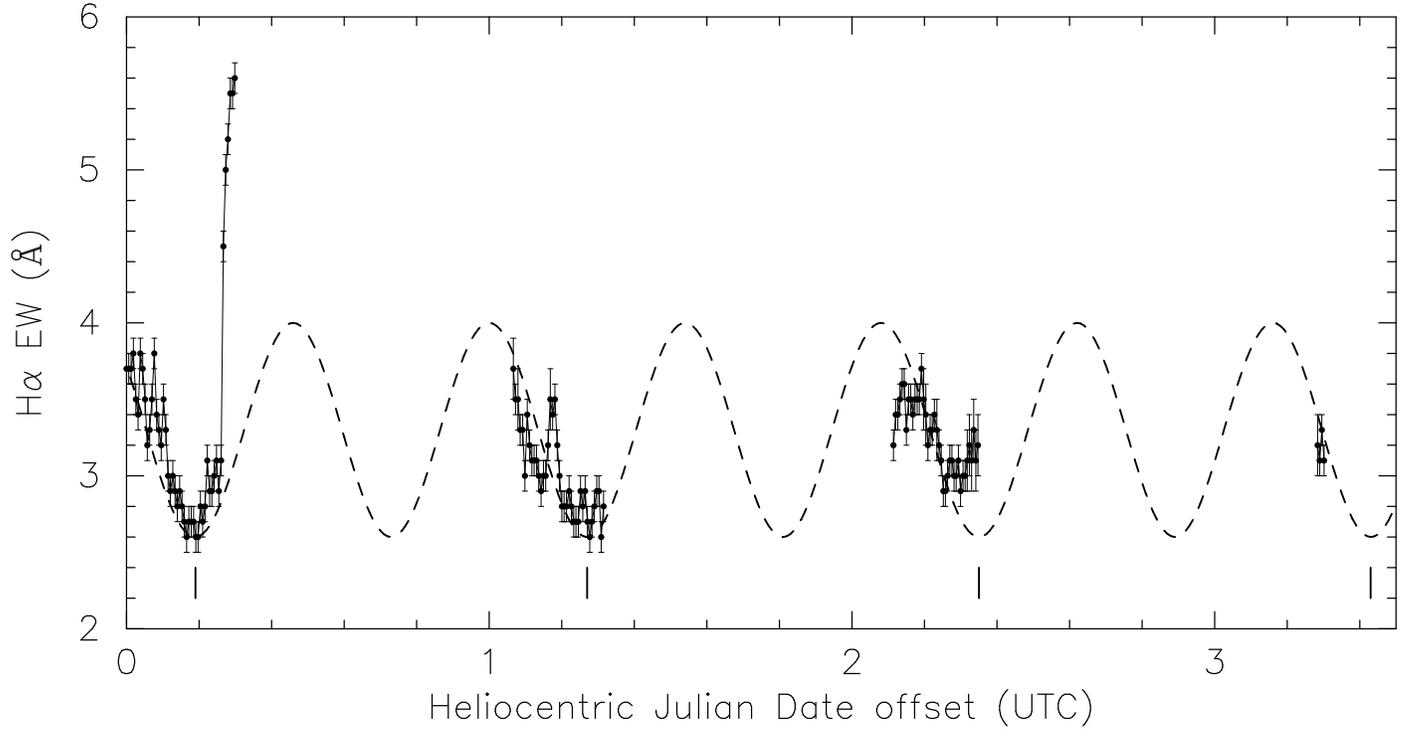}}
\caption{H$\alpha$ equivalent width light curve. The curve
appears likely sinusoidal, a sinusoidal curve (dashed-line) is also 
shown for a comparison. 
The first vertical tick at 0.19 in the HJD offset (HJD$-$2454549.79717)
position indicates
the bottom of the light curve obtained in the first night.
The remaining ticks are drawn with a distance of 
$2 \times P_{\rm rot}$ between two consecutive ticks where the star is observable, 
where $P_{\rm rot}=0.54$~days is the rotational period
determined from the Stokes $V$ profile set (see \S~3.2).
The observed bottoms of the light curve likely repeat after 2$P_{\rm rot}$,
suggesting that the H$\alpha$ emission is modulated by the stellar rotation.
Flaring events were observed in the first (strong) and second night (moderate). 
The decay and enhancement of H$\alpha$ emission were also
observed in the third night at around HJD offset$=$2.1 and 2.35, respectively. 
This is possibly due to a decrease (or an increase)
of the number of active regions. Changes in the
H$\alpha$ (also H$\beta$) line profile shape (e.g., smoothness) 
seen in our spectra support this possibility.
\label{f6}}
\end{figure}


\begin{thebibliography}{}

\bibitem[Audard et al.(2007)]{audard}
Audard, M., Osten, R. A., Brown, A., Briggs, K. R., G\"udel, M., 
Hodges-Kluck, E., \& Gizis, J. E. 2007, \aap, 471, L63

\bibitem[Baraffe et al.(1998)]{b98}
Baraffe, I., Chabrier, G., Allard, F., \& Hauschildt, P. H. 1998, \aap, 337, 403 

\bibitem[Berger(2002)]{berger02}
Berger, E. 2002, \apj, 572, 503 

\bibitem[Berger et al.(2008)]{berger08}
Berger, E., et al. 2008, \apj, 676, 1307

\bibitem[Berger et al.(2009)]{berger09}
Berger, E., et al. 2009, \apj, 695, 310

\bibitem[Brinkmann et al.(2000)]{brink}
Brinkmann, W., Laurent-Muehleisen, S. A., Voges, W., Siebert, J., Becker, R. H., 
Brotherton, M. S., White, R. L., \& Gregg, M. D. 2000, \aap, 356, 445

\bibitem[Brown \& Landstreet(1981)]{brown}
Brown, D. N., \& Landstreet, J. D. 1981, \apj, 246, 899

\bibitem[Brown et al.(1991)]{brown91}
Brown, S., Donati, J.-F., Rees, D., \& Semel, M. 1991, \aap, 250, 463

\bibitem[Browning(2008)]{browning}
Browning, M. K. 2008, \apj, 676, 1262

\bibitem[Cameron(1992)]{cameron92}
Cameron  A., 1992, in  Lecture Notes in Physics 397, Surface Inhomogeneities on Late-Type Stars,
ed. P. B. Byrne \& D. J. Mullan (Berlin: Springer-Verlag), 33

\bibitem[Chabrier \& Baraffe(1997)]{chabrier97}
Chabrier, G., \& Baraffe, I. 1997, \aap, 327, 1039

\bibitem[Chabrier \& K\"uker(2006)]{chabrier06}
Chabrier, G., \& K\"uker, M. 2006, \aap, 446, 1027

\bibitem[Chabrier et al.(2007)]{chabrier07}
Chabrier, G., Gallardo, J., \& Baraffe, I. 2007, \aap, 472, L17

\bibitem[Cram \& Giampapa(1987)]{cram87}
Cram, L. E., \& Giampapa, M. S. 1987, \apj, 323, 316

\bibitem[Delfosse et al.(1998)]{del98}
Delfosse, X., Forveille, T., Perrier, C., \& Mayor, M. 1998, \aap, 331, 581

\bibitem[Delfosse et al.(1999)]{del99}
Delfosse, X., Forveille, T., Beuzit, J.-L., Udry, S., Mayor, M., \& Perrier, C. 1999, \aap, 344, 897

\bibitem[Delfosse et al.(2000)]{delfosse00}
Delfosse, X., Forveille, T., S\'egransan, D., Beuzit, J.-L., Udry, S., Perrier, C.,
\& Mayor, M. 2000, \aap, 364, 217

\bibitem[Demory et al.(2009)]{demory}
Demory, B.-O., et al. 2009, \aap, in press (arXiv:0906.0602)

\bibitem[Dobler et al.(2006)]{dobler}
Dobler, W., Stix, M., \& Brandenburg, A. 2006, \apj, 638, 336

\bibitem[Donati et al.(1997a)]{donati97a}
Donati, J.-F., Semel. M., Carter, B. D., et al. 1997a, \mnras, 291, 658

\bibitem[Donati et al.(1997b)]{donati97b}
Donati, J.-F., \& Brown, S. F. 1997b, \aap, 326, 1135

\bibitem[Donati et al.(2001)]{donati01}
Donati, J.-F., Wade, G. A., Babel, J., Henrichs, H. F., 
de Jong, J. A., \& Harries, T. J. 2001, MNRAS, 326, 1265

\bibitem[Donati(2003)]{donati03}
Donati, J.-F. 2003, in ASP Conf. Ser. 307, Solar Polarization, 
ed. J. Trujillo-Bueno \& J. Sanchez Almeida (San Francisco: ASP), 41 

\bibitem[Donati et al.(2006a)]{donati06a}
Donati, J.-F., Forveille, T., Cameron, A. C., Barnes, J. R., 
Delfosse, X., Jardine, M. M., \& Valenti, J. A. 2006a, Science, 311, 633

\bibitem[Donati et al.(2006b)]{donati06b}
Donati, J.-F., et al. 2006b, MNRAS, 370, 629

\bibitem[Donati et al.(2008)]{donati08}
Donati, J.-F., et al. 2008, MNRAS, 390, 545

\bibitem[Durney et al.(1993)]{durney}
Durney, B. R., De Young, D. S., \& Roxburgh, I. W. 1993, Sol. Phys., 145, 207

\bibitem[Epchtein(1997)]{epchtein} Epchtein, N. 1997, 
in Garzon F., Epchtein N., Omont A., Burton B., Persi P., eds,
The impact of large scale near-IR sky surveys, Kluwer Academic Publishers, Dordrecht, p. 15

\bibitem[ESA(1997)]{esa}
ESA 1997, The Hipparcos and Tycho Catalogues, ESA SP-1200

\bibitem[Fleming et al.(1989)]{fleming89}
Fleming, T. A., Gioia, I. M., Maccacaro, T. 1989, \apj, 340, 1011

\bibitem[Fleming(1998)]{fleming98}
Fleming, T. A. 1998, \apj, 504, 461

\bibitem[G\"udel et al.(1993)]{gudel93}
G\"udel, M., Schmitt, J. H. M. M., Bookbinder, J. A., \& Fleming, T. A. 1993, \apj, 415, 236

\bibitem[Hallinan et al.(2008)]{hallinan08}
Hallinan, G., Antonova, A., Doyle, J. G., 
Bourke, S., Lane, C., \& Golden, A. 2008, \apj, 684, 644

\bibitem[Horne(1986)]{horne}
Horne, K. 1986, PASP, 98, 609

\bibitem[H\"unsch et al.(1999)]{huensch}
H\"unsch, M., Schmitt, J. H. M. M., Sterzik, M. F., \& Voges, W. 1999, \aaps, 135, 319

\bibitem[Jardine et al.(1999)]{jardine99}
Jardine, M., Barnes, J. R., Donati, J.-F., \& Collier Cameron, A. 1999, MNRAS, 305, 35

\bibitem[Johns-Krull \& Valenti(1996)]{jk96}
Johns-Krull, C. M., \& Valenti, J. A. 1996, \apj, 459, L95

\bibitem[Johns-Krull \& Valenti(2000)]{jk00} 
Johns-Krull, C.~M., \& Valenti, J.~A.\ 2000, Stellar Clusters and 
Associations: Convection, Rotation, and Dynamos, 198, 371 

\bibitem[Johns-Krull et al.(1999)]{jk99} 
Johns-Krull, C.~M., Valenti, J.~A., \& Koresko, C.\ 1999, \apj, 516, 900

\bibitem[K\"uker \& R\"udiger(1997)]{kuker97}
K\"uker, M., \& R\"udiger, G. 1997, \aap, 328, 253

\bibitem[K\"uker \& R\"udiger(1999)]{kuker99}
K\"uker, M., \& R\"udiger, G. 1999, \aap, 346, 922

\bibitem[Kurucz(1993)]{kurucz}
Kurucz, R. L. 1993, CDROM \# 13 (ATLAS9 atmospheric models) and \# 18 (ATLAS9 and SYNTHE routines,
spectral line database).

\bibitem[Landolfi \& Landi Degl'Innocenti(1982)]{lando} 
Landolfi, M., \& Landi Degl'Innocenti, E. 1982, Sol. Phys., 78, 355

\bibitem[Leggett(1992)]{leggett} 
Leggett, S. K. 1992, \apjs, 82, 531

\bibitem[Leto et al.(2000)]{leto}
Leto, G., Pagano, I., Linsky, J. L., Rodon\`o, M., \& Umana, G. 2000, \aap, 359, 1035

\bibitem[Liebert et al.(2003)]{liebert}
Liebert, J., Kirkpatrick, J. D., Cruz, K. L., Reid, I. N., Burgasser, A., Tinney, C. G.,
\& Gizis, J. E. 2003, \aj, 125, 343

\bibitem[Lim(1993)]{lim}
Lim, J. 1992, \apj, 405, L33

\bibitem[Marcy(1982)]{marcy82}
Marcy, G. W. 1982, PASP, 94, 989

\bibitem[Mart\'{\i}n et al.(1999)]{martin99}
Mart\'{\i}n, E. L., Delfosse, X., Basri, G., Goldman, B., Forveille, T., 
\& Zapatero Osorio, M. R. 1999, \aj, 118, 2466

\bibitem[Mart\'{\i}nez Gonz\'alez et al.(2008)]{gonzalez}
Mart\'{\i}nez Gonz\'alez, M. J., Asensio Ramos, A., Carroll, T. A., 
Kopf, M., Ram\'{\i}rez V\'elez, J. C., \& Semel, M. 2008, \aap, 486, 637

\bibitem[Mochnacki et al.(2002)]{moch}
Mochnacki, S. W., et al. 2002, \aj, 124, 2868

\bibitem[Morales et al.(2008)]{morales}
Morales, J. C., Ribas, I., \& Jordi, C. 2008, \aap, 478, 507

\bibitem[Morin et al.(2008)]{morin08}
Morin, J., et al. 2008, MNRAS, 390, 567

\bibitem[Mullan \& MacDonald(2001)]{mullan}
Mullan, D. J., \& MacDonald, J. 2001, \apj, 559, 353

\bibitem[Noyes et al.(1984)]{noyes}
Noyes, R. W., Hartmann, L. W., Baliunas, S. L., Duncan, D. K., \& Vaughan, A. H. 1984, \apj, 279, 763

\bibitem[Oppenheimer et al.(2001)]{oppen}
Oppenheimer, B. R., Golimowski, D. A., Kulkarni, S. R., Matthews, K.,
Nakajima, T., Creech-Eakman, M., \& Durrance, S. T. 2001, \aj, 121, 2189

\bibitem[Orlov et al.(1995)]{orlov}
Orlov, V. V., Panchenko, I. E., Rastorguev, A. S., \& Yatsevich, A. V. 1995, AZh, 72, 495

\bibitem[Osten et al.(2009)]{osten09}
Osten, R. A., Phan-Bao, N., Hawley, S. L., Reid, I. N., 
\& Ojha, R. 2009, \apj, in press (arXiv:0905.4197)

\bibitem[Parker(1975)]{parker75}
Parker, E. N. 1975, \apj, 198, 205

\bibitem[Phan-Bao et al.(2006)]{pb06}
Phan-Bao, N., Mart\'{\i}n, E. L., Donati, J.-F., \& Lim, J. 2006, \apj, 646, L73

\bibitem[Phan-Bao et al.(2007)]{pb07}
Phan-Bao, N., Osten, R. A., Lim, J., Mart\'{\i}n, E. L., \& Ho, P. T. P. 2007, \apj, 658, 553

\bibitem[Rao \& Singh(1990)]{rao}
Rao, A. R., \& Singh, K. P., 1990, \apj, 352, 303

\bibitem[Reid et al.(1995)]{reid95}
Reid, I. N., Hawley, S. L., \& Gizis, J. E. 1995, \aj, 110, 1838

\bibitem[Reiners \& Basri(2007)]{rn07}
Reiners, A., \& Basri, G. 2007, \apj, 656, 1121  

\bibitem[Ribas(2006)]{ribas}
Ribas, I. 2006, Ap\&SS, 304, 89

\bibitem[Roberts \& Stix(1972)]{roberts72}
Roberts, P. H., \& Stix, M. 1972, \aap, 18, 453

\bibitem[Robrade et al.(2009)]{robrade}
Robrade, J., \& Schmitt, J. H. M. M. 2009, \aap, in press (arXiv:0901.3027)

\bibitem[R\"udiger \& Elstner(1994)]{rudiger}
R\"udiger, G., \& Elstner, D. 1994, \aap, 281, 46

\bibitem[Saar(1988)]{saar88}
Saar, S. H. 1988, \apj, 324, 441

\bibitem[Saar(1994)]{saar94}
Saar, S. H. 1994, in IAU Symp. 154, Infrared Solar Physics, ed. D. M. Rabin et al.
(Dordrecht: Kluwer), 493

\bibitem[Schmidt et al.(2007)]{schmidt07}
Schmidt, S. J., Cruz, K. L., Bongiorno, B. J., Liebert, J., \& Reid, I. N. 2007, \aj, 133, 2258 

\bibitem[Sch\"ussler(1975)]{schu}
Sch\"ussler, M. 1975, \aap, 38, 263

\bibitem[Segransan et. al.(2003)]{segransan}
S\'egransan, D., Kervella, P., Forveille, T., \& Queloz, D. 2003, \aap, 397, L5

\bibitem[Semel et al.(1989)]{semel89}
Semel, M. 1989, \aap, 225, 456

\bibitem[Semel et al.(2008)]{semel}
Semel, M., Ramirez V\'elez, J. C., Stift, M. J., 
Mart\'{\i}nez Gonz\'alez, M. J., L\'opez Ariste, A. \& Leone, F. 2008, arXiv:0810.3543 

\bibitem[Singh et al.(1999)]{singh}
Singh, K. P., Drake, S. A., Gotthelf, E. V., \& White, N. E. 1999, \apj, 512, 874

\bibitem[Skilling \& Bryan(1984)]{ski}
Skilling, J., \& Bryan, R. K. 1984, MNRAS, 211, 111

\bibitem[Solanki(2009)]{so09} Solanki, S.~K.\ 2009, 
Astronomical Society of the Pacific Conference Series, 405, 135

\bibitem[Song et al.(2002)]{song}
Song, I., Bessell, M., \& Zuckerman, B. 2002, \apj, 581, L43

\bibitem[Stauffer \& Hartmann(1986)]{stauffer86}
Stauffer, J. R., \& Hartmann, L. W. 1986, \apjs, 61, 531

\bibitem[Stelzer et al.(2006)]{stelzer06}
Stelzer, B., Schmitt, J. H. M. M., Micela, G., \& Liefke, C. 2006, \aap, 460, 35

\bibitem[Tinney et al.(1993)]{tinney}
Tinney, C. G., Mould, J. R., \& Reid, I. N. 1993, \aj, 105, 1045

\bibitem[Vaiana et al.(1981)]{vaiana}
Vaiana, G. S., et al. 1981, \apj, 244, 163

\bibitem[van Altena(1995)]{van95}
van Altena,  W. F., Lee,  J. T., \& 
Hoffleit,  E. D. 1995, The General Catalogue of Trigonometric Stellar Parallaxes, 
4th edn. Yale Univ. Observatory,  New Haven, CT  

\bibitem[Vogt(1980)]{vogt}
Vogt, S. S. 1980, \apj, 240, 567

\bibitem[Vogt et al.(1987)]{vogt87}
Vogt, S. S., Penrod, G. D., \& Hatzes, A. P. 1987, \apj, 321, 496

\bibitem[Wade et al.(2000)]{wade}
Wade, G. A., Donati, J.-F., Landstreet, J. D., \& Shorlin, S. L. S. 2000, MNRAS, 313, 851

\bibitem[XMM-SSC(2008)]{xmm}
XMM-SSC 2008, The XMM-Newton 2nd Incremental Source Catalogue (2XMMi), 
(Leicester, UK: XMM-SSC)

\bibitem[Young et al.(1984)]{young84}
Young, A., Skumanich, A., \& Harlan, E. 1984, \apj, 282, 683

\bibitem[Young et al.(1990)]{young90}
Young, A., Skumanich, A., MacGregor, K. B., Temple, S. 1990, \apj, 349, 608

\bibitem[Zuckerman \& Song(2004)]{zuckerman}
Zuckerman, B., \& Song, I. 2004, ARA\&A, 42, 685

\end{thebibliography}
\end{document}